\documentclass[showpacs,aps,pre,amsfonts,amsmath,amssymb,nofootinbib,10pt,twocolumn]{revtex4-1}
\pdfoutput=1
\usepackage{graphicx}
\usepackage[colorlinks,linkcolor=blue,citecolor=blue,urlcolor=blue]{hyperref}
\usepackage{appendix}
\graphicspath{{figures/},.}
\renewcommand{\S}{\mathcal{S}}
\begin{document}
\title{Imitation with incomplete information in 2$\times$2 games}
\author{Mathis Antony}
\email{mathis.antony@gmail.com}
\author{Degang Wu}
\email{samuelandjw@gmail.com}
\author{K Y Szeto}
\email[corresponding author, ]{phszeto@ust.hk}
\affiliation{Department of Physics \\ Hong Kong University of Science and Technology\\Clear Water Bay, Hong Kong}
\begin{abstract}
Evolutionary game theory has been an important tool for describing economic and social behaviour for decades. Approximate mean value equations describing the time evolution of strategy concentrations can be derived from the players' microscopic update rules. We show that they can be generalized to a learning process. As an example, we compare a restricted imitation process, in which unused parts of the role model's meta-strategy are hidden from the imitator, with the widely used imitation rule that allows the imitator to adopt the entire meta-strategy of the role model. This change in imitation behaviour greatly affects dynamics and stationary states in the iterated prisoner dilemma. Particularly we find Grim Trigger to be a more successful strategy than Tit-For-Tat especially in the weak selection regime.
\end{abstract}
\maketitle
\section{Introduction \label{sec:intro}}
In evolutionary game theory \cite{Smith82,Axelrod84,Hofbauer98,Nowak06b} the success of selfish individuals (players) is determined by their interaction with other players. The most extensively studied evolutionary two player game is the \emph{iterated} (or \emph{repeated}) \emph{Prisoner Dilemma} (IPD) \cite{Poundstone92}. The rules of the Prisoner Dilemma game are brief and simple yet the parallels that can be drawn to human behaviour are manifold. The evolutionary aspect of the IPD is commonly introduced by imitative behaviour of the players or a reproduction process. Here we investigate a restriction on the players' ability to imitate, in which imitators can only use information obtained by interacting with the role model. We incorporate this adjustment in the \emph{approximate mean value equation} \cite{Szabo07} that describes the time evolution of strategy concentrations. Particularly we consider the simplest case of a one step memory where players remember what happened in the last encounter of the game. We show that while populations eventually reach a cooperative equilibrium, GT is the dominant strategy and the only surviving strategy in the weak-selection regime with partial imitation.

The \emph{Prisoner Dilemma} and other symmetric $2\times 2$ games (such as the \emph{Chicken} or \emph{Snowdrift} game and the \emph{Stag-Hunt} game) are defined by the following set of rules. When two players play a game each of them chooses to \textit{cooperate} ($C$) or \textit{defect} ($D$). Based on these two choices the two players are attributed a payoff. A cooperating player scores $R$ ($S$) if his opponent cooperates (defects) and a defecting player scores $T$ ($P$) if his opponent cooperates (defects). Usually $R$ is called the \textit{Reward for cooperation}, $S$ the \textit{Sucker's payoff}, $T$ the \textit{Temptation} to defect and $P$ the \textit{Punishment} for mutual defection. The Prisoner Dilemma game imposes the following restrictions on the payoff parameters: $T>R>P>S$ and $2R>T+P$ to prevent collusion. In the restrictions on the payoff parameters roots the tragedy of the Prisoner Dilemma. The strategy with the highest expectation for a player is $D$, while the strategy that yields the highest payoff for the population is $C$. The PD is a thus \textit{non-zero sum} game as one player's loss does not equal his opponent's gain. As defection dominates cooperation, defection is the rational strategy to choose for any player and mutual defection is the only \emph{Nash equilibrium}. Cooperation is however a widespread phenomenon in nature and the study of the emergence of cooperation in a population of selfish individuals has been one of the key objectives in evolutionary game theory over the past few decades.

Nowak and his collaborators have contributed enormously to this topic \cite{Nowak92,Nowak06b,Nowak94,Ohtsuki06,Nowak93,Nowak04,Pacheco08,Ohtsuki08} and particularly they found five rules of cooperation \cite{Nowak06}. Among them, direct \cite{Pacheco08,Nowak04b,Nowak06} and indirect reciprocity \cite{Nowak05,Nowak06} may lead to cooperative behaviour. In complex networks, cooperators can support each other in more than one dimension, as for example in \cite{Zimmermann05,Wu07,Nowak93,Pacheco08,Szabo05,Ohtsuki06,Ohtsuki08}. Qin et al. observe the emergence of cooperation among players with the ability to recall their payoff over several generations\cite{qin08}. If players may recall what happened in the last game they may employ the famous \emph{Tit-For-Tat} (TFT), \emph{Pavlov} and \emph{Grim Trigger} (GT) strategies. More sophisticated players that make decisions based on their own and their opponent's moves in recent games are used in \cite{Baek08,Haider06,Brunauer07}. We follow a similar approach in this paper. Our players decide how to play based only on the outcome of the most recent game.

In evolutionary game theory it is common to define a measure for fitness that is a monotonously increasing function of the payoff and to assume that higher fitness leads to higher reproduction rates. \emph{Approximate mean value} or \emph{replicator} equations may be used to predict the fate of different strategies for large, well mixed populations. Players with tendency to imitate promising strategies are commonly used at the microscopic level. However if strategies more complex than simply \emph{cooperate} or \emph{defect} -- so called \emph{meta-strategies} -- are used, imitating may turn out to be tricky. Imagine the situation of  a novice chess player trying to learn from more experienced participants in a chess tournament. The novice may gradually improve his game by incorporating the behaviour of better participants in his own strategy, but he will not be able to extract and learn the complete strategy of other players immediately. In the same spirit, the partial imitation learning process involves imitation of the exposed part of the role model's strategy. A detailed example is illustrated in section \ref{sec:pir}. In the context of the PD game, we introduce the approximate mean value equation for the \emph{partial Imitation Rule} (pIR) and compare it with the the \emph{traditional Imitation Rule} (tIR), where the complete strategy can be copied by the learner.

The rest of the present paper is structured as follows: in section \ref{sec:methods} we describe our methods, results are presented in section \ref{sec:results} and discussed in section \ref{sec:discussion}. We draw our conclusion in section \ref{sec:conclusion}.
\section{Methods \label{sec:methods} }
In section \ref{sec:memory} we give a precise description of memory and define the possible one-step strategies that appear in this paper. The partial Imitation Rule (pIR) is described in detail in section \ref{sec:pir} and the arising equations describing macroscopic dynamics are discussed in \ref{sec:macro}.
\subsection{Memory \label{sec:memory}}
The ensemble of possible strategies for players with $n$ steps memory is denoted as $M_n$. As mentioned previously we allow our agents to play moves based on their own previous move and the move of their opponent. Therefore we need an encoding scheme for $M_1$ strategies. As every player has two choices for each move ($D$ or $C$) there are $4$ possible outcomes ($DD$, $DC$, $CD$ and $CC$) with payoffs $P,\,T,\,S$ and $R$ every time the game is played. Thus we need $4$ responses $\S_P$, $\S_T$, $\S_S$ and $\S_R$ for the  $DD$, $DC$, $CD$ and $CC$ histories of the last game respectively. The agents also need to know how to start playing if there is no history. We add an additional first move $\S_0$. Adding up to a total of $5$ moves for a one-step memory strategy. A strategy in $M_1$ is thus denoted as $\S_0|\S_P\S_T\S_S\S_R$ where $\S_0$ is the first move and $\S_P$, $\S_T$, $\S_S$ and $\S_R$  are the moves that follow $DD$, $DC$, $CD$ and $CC$ histories respectively. Thus there are $|M_1|=2^5=32$ possible strategies as there are two choices for each $\S_i$, either $C$ or $D$. In table \ref{tab:strat} this scheme is illustrated along with three famous strategies Grim-Trigger, Tit-For-Tat and Pavlov and the groups of $4$ strategies that always defect (cooperate) in practice.
\begin{table}[ht]
\centering
\caption{Strategy sequences in $M_1$\label{tab:strat}. Omitted fields may be either C or D.}
\begin{ruledtabular}
\begin{tabular}{lccccc}
History: & - & DD & DC & CD & CC  \\
\hline
Move: & $\S_0$ & $\S_P$ & $\S_T$ & $\S_S$ & $\S_R$ \\
\hline
GT & C & D & D & D & C \\
\hline
TFT & C & D & C & D & C  \\
\hline
Pavlov & C & C & D & D & C \\
\hline
always defect & D & D & D &  &  \\
\hline
always cooperate & C &  &  & C & C \\ 
\hline 
nice & C &  &  &  & C \\
\hline
retaliating &  &  &  & D & 
\end{tabular}
\end{ruledtabular}
\end{table}
The aforementioned encoding scheme is easily generalized to $M_n$. A treatment of players with two-step, three-step and even longer memory can be found in \cite{Baek08,Haider06,Brunauer07}. Note that the total number of possible strategies $|M_n|$ increases exponentially with $n$. 
\subsection{partial Imitation Rule (pIR) \label{sec:pir}}
\begin{figure}[ht]
\centering
\includegraphics[width=0.5\linewidth]{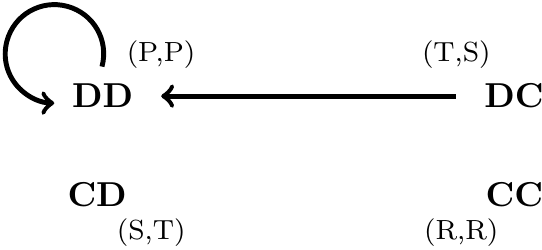}
\caption{Transition graph between the D$|$DDDD strategist Alice and the TFT (C$|$DCDC) strategist Bob. In parentheses the payoff of Alice, Bob from the corresponding state. The payoff from the recurrent state is $P$ for both strategies.}
\label{fig:tft-cdddd}
\end{figure}

As mentioned earlier \textit{pIR} should be a reasonable restriction of the agents' abilities to imitate. We explain in more detail using a concrete example. Consider Alice using strategy D$|$DDDD $= \S_0^A|\S_P^A\S_T^A\S_S^A\S_R^A$ playing against Bob, who is himself a \textit{TFT} (C$|$DCDC $= \S_0^B|\S_P^B\S_T^B\S_S^B\S_R^B$) strategist. The transition graph for this encounter is shown in figure \ref{fig:tft-cdddd}.
From Alice's point of view the first outcome is $\S_0^A\S_0^B=DC$, hence she uses $\S_T^A=D$ for her next move. Similarly Bob uses $\S_S^B = D$. The outcome of the second game is thus $\S_T^A\S_S^B=DD$.  Both players then use their $\S_P$ move $D$ for the third game. All subsequent outcomes are therefore $\S_P^A\S_P^B=DD$ and Alice's and Bob's recurrent state payoff, i.e. the average payoff per game in the limit where infinitely many games are played between the two, is therefore $P$. In summary, Alice has plays $\S_0^A\,\S_T^A\,\S_P^A\,\S_P^A\,...$ and Bob plays $\S_0^A\,\S_S^A\,\S_P^A\,\S_P^A\,...$. In the framework of \textit{pIR} Alice now imitates Bob. As she has only witnessed him play his $\S_0$, $\S_P$ and $\S_S$ move she will only adopt these moves from Bob's strategy. Her strategy will be changed in the imitation process as
\begin{equation}
\begin{array}{cccc}
\S_0^A|\S_P^A\S_T^A\S_S^A\S_R^A & \longrightarrow &  \S_0^B|\S_P^B\S_T^A\S_S^B\S_R^A & ,\quad \text{or} \\ \noalign{\smallskip} 
\text{D$|$DDDD} & \xrightarrow{\text{TFT}} & \text{C$|$DDDD} &
\end{array} \nonumber
\end{equation}
where $i\xrightarrow{k}j$ means that strategy $i$ turns into strategy $j$ by imitating strategy $k$. The next round Alice will thus be playing C$|$DDDD. We see that the result of such an imitation process may differ from a complete imitation of the role model's strategy (in this case Alice would simply become a TFT player) even in the simple case of players with one-step memory. To avoid confusion we refer to the rule that allows the imitator to copy the entire strategy of the role model \textit{traditional Imitation Rule} (tIR). The example illustrates how pIR strips the imitators from a supernatural ability to ``mind read" the role models and thereby extract hidden parts of their strategy. Note that in the case where players do not have memory, i.e. are simply cooperators or defectors, there is no need for a distinction between tIR and pIR. If agents with different memory lengths are interacting we need to specify how an agent Alice with a $n$-step memory imitates an agent Bob with a longer $m$-step memory if the sequence of Bob's moves witnessed by Alice cannot be mapped on an $n$-step memory strategy. As only one-step memory players are used here we do not address the issue this paper.

In order for our players to effectively make use of their memory and strategies, players need to play repeatedly against their opponents. We denote the number of games played per encounter of two players as $f$. This number affects the performance of different strategies and fate of the population in a complex way and this discussion is beyond the scope of this paper. If we assume that two players always play many games against each other before they switch to another opponent, then the players will mostly find themselves in recurrent states of the transition graph. In the limit where $f\rightarrow \infty$ the average payoff per game played is the average payoff scored in the recurrent states of the transition graphs. Thus in our large, well mixed population the payoff of a player playing strategy $i$ is well approximated by
\begin{equation}
U_i = \sum\limits_{j=1}^{|M_1|} \rho_j U_{ij} \, .
\end{equation}
where $\rho_j$ is the (number) density, concentration or fraction of $j$-strategists in the population and $U_{ij}$ is the average recurrent state payoff obtained by an $i$-strategist playing against a $j$-strategist. We fix our PD payoff parameters to a set of commonly used values that offer high temptation to defecting players: $T=5,\, R=3,\, P=1,\, S=0$. The same set was also used in Axelrod's famous computer tournaments \cite{Axelrod84}.
\subsection{Macroscopic Dynamics\label{sec:macro}}
At the macroscopic level we are interested in players that do not necessarily make perfect decisions when imitating other players. To account for these irrational decisions we assign a certain probability to every possible imitation process depending on the payoff difference between the imitator and the role model. If player $i$ has been determined as a possible imitator of player $j$ the imitation (with tIR or pIR) occurs with a probability given by a monotonically increasing smoothing function $g(\Delta U)$, where $\Delta U=U_j-U_i$ is the payoff difference between player $i$ and $j$. This translates into the fact that the more successful players are, the more likely they are to be imitated. With our choice of smoothing function $g$ the imitation probability is given by
\begin{eqnarray}
P(i \textrm{ imitates } j) &=& g(\Delta U) =  \frac{1}{1+\exp \left( \frac{-\Delta U}{K} \right)} \nonumber \\ &=& \frac{1}{1+\exp \left( \frac{U_i-U_j}{K} \right)} \label{eq:g}
\end{eqnarray}
where $K>0$ is a temperature-like noise factor controlling the extent of irrationality among the players\cite{Szabo07}.

By using this combination of pIR and smoothing function $g$ defined above associated to the payoff of the players we basically make assumptions about the availability and reliability of information. The details about encounters between two players Alice and Bob (i.e. the exact moves played during the encounter) are known to Alice and Bob only. We also assume that Alice and Bob do not make any mistakes when memorizing moves and when using their strategy to play. However the information about the total payoff of players is available globally to all players but associated with an uncertainty whose extent is controlled by the noise factor $K$. In this way the players' first hand information is reliable but severely limited if the imitation process is based on pIR and information about the wealth of the players is available globally but this information is not perfectly reliable.

It is possible to determine macroscopic dynamics from the microscopic update rules for both tIR and pIR \cite{Szabo07} (see appendix \ref{sec:app}).
The \textit{approximate mean value equations} for tIR is
\begin{equation}
\frac{d \rho_i^{\textrm{tIR}}}{dt}= \rho_i \sum\limits_{j}  \rho_j  \left[ g(U_i-U_j)- g(U_j-U_i)  \right] \label{eq:mean_tir}
\end{equation}
where all the sums are carried out over all considered strategies. If as in the case of pIR the imitator may adopt a strategy that is different from the role models strategy the approximate mean value equation is
\begin{eqnarray}
\frac{d \rho_i^{\textrm{pIR}}}{dt}&=& \phantom{-}\sum\limits_{j} \rho_j \sum\limits_{k}   \rho_k
 g(U_j-U_k)p (k,j,i) \nonumber \\ && -  \rho_i \sum\limits_j \rho_j g(U_j-U_i)  \,. 
\end{eqnarray}
where
\begin{equation}
p : M \times M \times M  \rightarrow  [0, 1] 
\end{equation}
is a function whose value $p(k,j,i)$ is the probability that $k$-strategist will become an $i$-strategist by imitating a $j$-strategist and $M$ is the space of all allowed strategies. Because the player must use a strategy after the imitation process we have the restriction
\begin{equation}
\sum_i p(k,j,i) = 1 \quad \forall k,j \in M \, .
\end{equation}
For our model of partial imitation, this probability is reduced to a simple form
\begin{eqnarray}
p_{\textrm{pIR}} &:& M \times M \times M  \rightarrow  \lbrace 0, 1\rbrace \\
p_{\textrm{pIR}}(k,j,i) &=& \left \lbrace \begin{array}{l}1\quad\textrm{if $k$-strategist imitating} \\ \textrm{\phantom {1}\quad$j$-strategist becomes}\\ \textrm{\phantom {1}\quad $i$-strategist} \\ 0\quad\textrm{otherwise} \end{array} \right. \label{eq:ppir} .
\end{eqnarray}
The approximate mean value equation for pIR is thus
\begin{eqnarray}
\frac{d \rho_i^{\textrm{pIR}}}{dt}&=& \phantom{-}\sum\limits_{j} \rho_j \sum\limits_{k}   \rho_k
 g(U_j-U_k)p_{\textrm{pIR}} (k,j,i) \nonumber \\ && -  \rho_i \sum\limits_j \rho_j g(U_j-U_i)  \,. \label{eq:mean_pir}
\end{eqnarray}
\section{Results \label{sec:results}}
The dynamics of the two imitation rules are compared in section \ref{sec:dynamics} and the equilibrium strategy fractions presented in section \ref{sec:equilibrium}.
\subsection{Dynamics\label{sec:dynamics}}
\begin{figure}[ht]
\centering
\includegraphics[width=\linewidth]{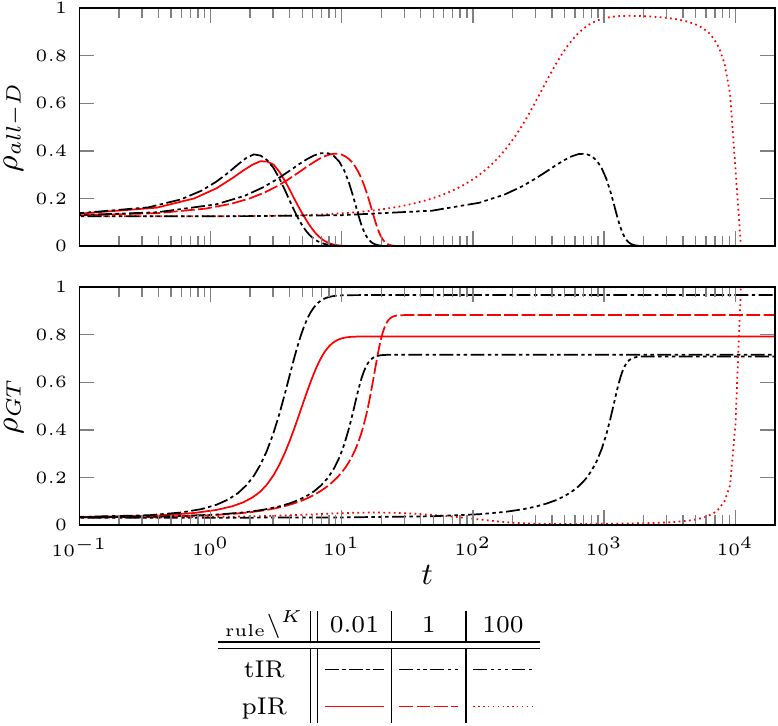}
\caption{Strategy concentrations $\rho$ as function of time $t$ for different noise factors $K$. Results from numerical integration of the approximate mean value equations for tIR (\ref{eq:mean_tir}) and pIR (\ref{eq:mean_pir}) with initial condition $\rho_i(t=0) = \frac{1}{|M_1|}$ for $i=1,2...,n$ where $|M_1|=32$ is the number of strategies in $M_1$.}
\label{fig:evo}
\end{figure}

We would like to observe the dynamics of the two imitation rules. Upon examining the smoothing function \eqref{eq:g} we distinguish three different cases. In the low noise limit we have rational players as $\lim\limits_{K \to\,0} g(\Delta U) = \Theta \left( \Delta U \right)$ where $\Theta \left( \cdot \right)$ is the heaviside step function. In high noise limit we have random drift\footnote{Note that if pIR is used then this random drift is possible only between certain strategies.} because $\lim\limits_{K \to \infty} g(\Delta U) = \frac{1}{2}$. In figure  \ref{fig:evo} the concentration of a selection of important strategies is given in three cases of low, medium and high noise.
We refer to the cumulative concentration of always defecting strategies (see table \ref{tab:strat}) as $\rho_{all-D}$. When tIR is used $\rho_{all-D}$ follows a similar evolution for all three noise factors. The always defecting strategies die out after their fraction increases initially to about 0.4. With more noise this process takes more time. If pIR is used we observe a similar evolution for low and medium noise. As the noise increases, however, we find that the population is temporarily dominated by the always defecting strategies before they die out eventually. By observing the fraction of GT players we notice a simultaneous extinction of all-D strategies and a rise of the fraction of GT players followed by an equilibrium state, dominated by the GT strategy.
\subsection{Equilibrium strategy distribution\label{sec:equilibrium}}
\begin{figure}
\centering
\includegraphics[width=\linewidth]{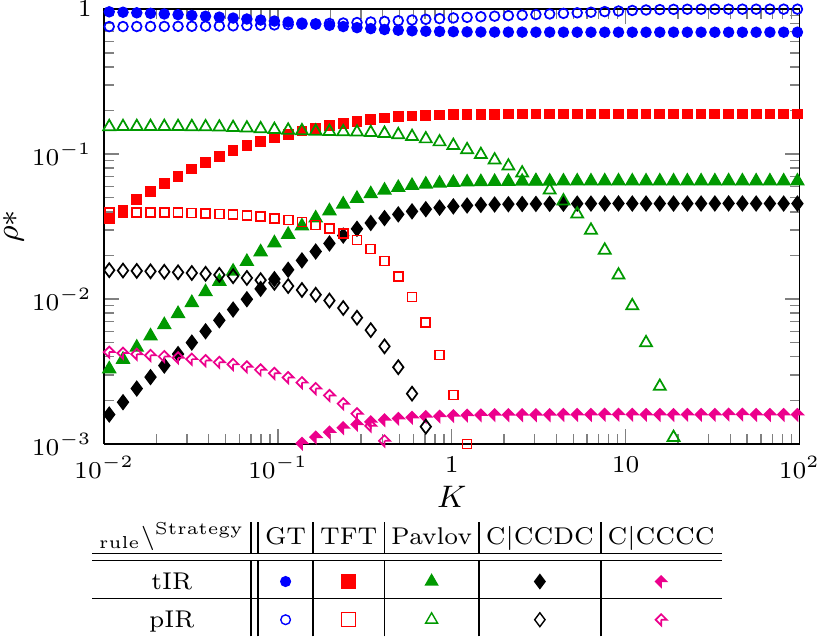}
\caption{Selection of non-zero equilibrium fractions as function of noise factor $K$. Results from numerical integration of the \textit{approximate mean value equations} for tIR, equation \ref{eq:mean_tir} and pIR, equation \ref{eq:mean_pir} with initial condition $\rho_i(t=0) = \frac{1}{|M_1|}$ for $i=1,2...,|M_1|$ where $|M_1|=32$ is the number of strategies in $M_1$.}
\label{fig:eq}
\end{figure}
The results from numerical integration of the \textit{approximate mean value equations} equation \ref{eq:mean_tir} and \ref{eq:mean_pir} suggest that if initially all the strategies are present in equal fractions, the system is  likely to reach a stationary equilibrium. We can see in figure \ref{fig:evo} that the equilibrium fraction of GT varies with the noise factor $K$. Figure \ref{fig:eq} shows the equilibrium fractions of all  nice and retaliating strategies and the C$|$CCCC strategy as function of noise factor $K$ for both imitation rules\footnote{With pIR and low noise a few other strategies have small non-zero equilibrium densities, namely C$|$DDCC, C$|$DCCC and C$|$CDCC. With tIR these three other always cooperating strategies have the same equilibrium fractions as the C$|$CCCC strategy.}. We denote the equilibrium fraction of strategy X as $\rho^{*}_\textrm X$. If we only refer to the equilibrium fraction under traditional (partial) imitation we add a tIR (pIR) superscript: $\rho^{*\textrm{tIR}}_\textrm X$ ($\rho^{*\textrm{pIR}}_\textrm X$).

GT is the most abundant strategy at equilibrium for both imitation rules and over the whole reasonable range of the noise factor $K$. For traditional imitation we notice that the lower the noise factor $K$, the higher the equilibrium fraction of GT $\rho^*_{\textrm{GT}}$ and for moderate to high noise factor, the equilibrium fractions are independent of the noise factor. These two observations are reversed for partial imitation, i.e. for pIR, the higher the noise factor $K$ the higher $\rho^*_{\textrm{GT}}$ and at low temperatures the equilibrium fractions are independent of $K$. We notice further that for traditional (partial) imitation GT is the only dominating strategy and $\rho^*_{\textrm{GT}}$ is very close to $1$ if the noise factor is small (high). With traditional imitation the equilibrium fractions rank independent of the noise factor as $\rho^{*\textrm{tIR}}_{\textrm{GT}} > \rho^{*\textrm{tIR}}_{\textrm{TFT}} > \rho^{*\textrm{tIR}}_{\textrm{Pavlov}} > \rho^{*\textrm{tIR}}_{\textrm C|\textrm{CCDC}} > \rho^{*\textrm{tIR}}_{\textrm C|\textrm{CCCC}}$.
For partial imitation this is no longer true. Pavlov is now more abundant than TFT at equilibrium. 
\section{Discussion \label{sec:discussion}}
\begin{figure}[ht]
\centering
\includegraphics[width=\linewidth]{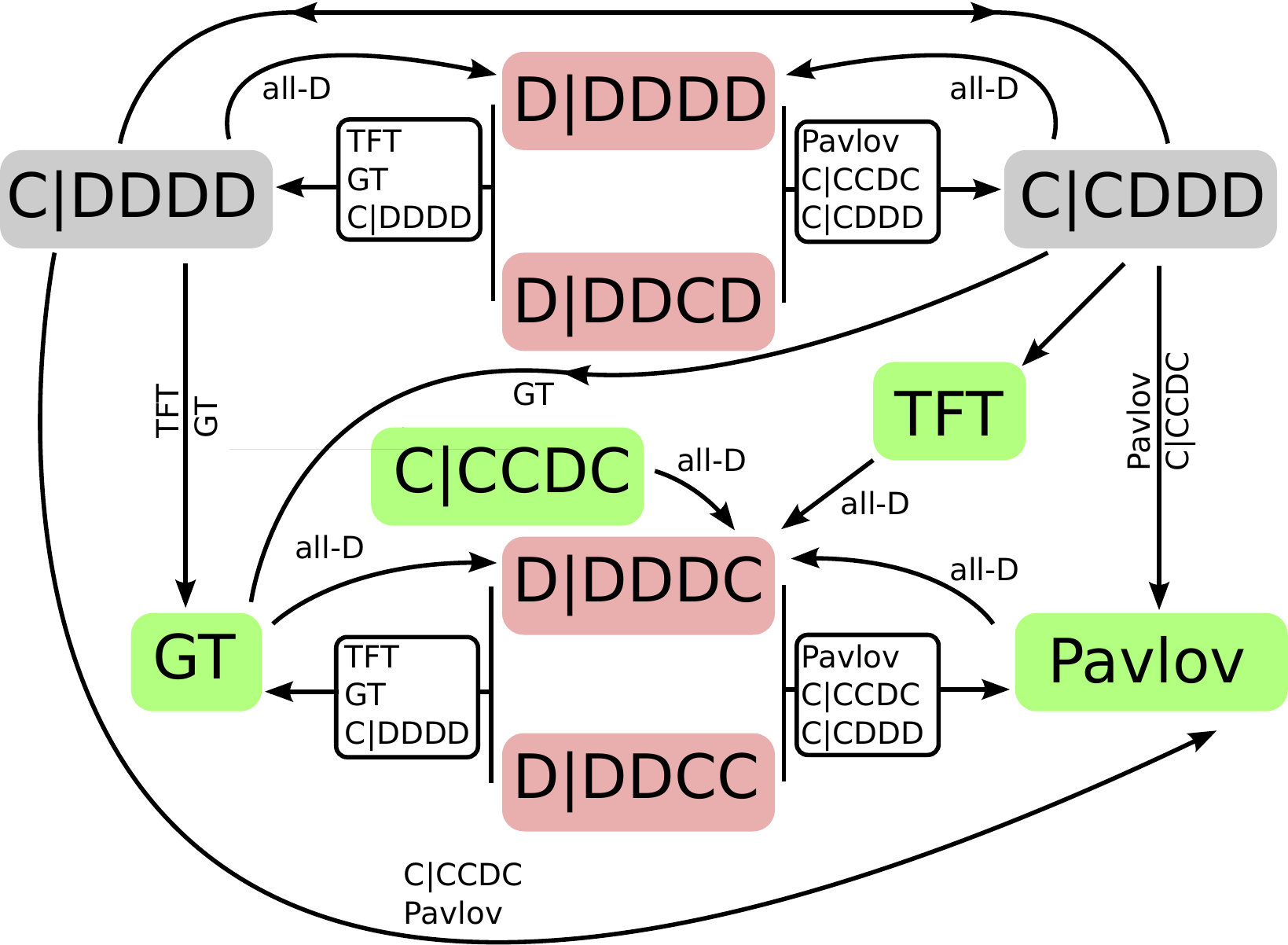}
\caption{Flow chart of important strategies under pIR. An arrow A$\xrightarrow{^{\;\,\textrm{C}}_{\textrm{(D)}}}$B is drawn if an A-strategist adapts strategy B when imitating a C-strategist (or a D-strategist). For simplicity we write A$\xrightarrow{\textrm B}$B as A$\rightarrow$B. If A is nice and retaliating strategy and B is one of the two transitional strategies (C$|$DDDD or C$|$CDDD) we also have A$\rightarrow$B. These arrows are omitted to avoid an overly crowded figure. The term all-D is used to denote all four always defecting strategies.}
\label{fig:chart}
\end{figure}

With pIR there are $|M_1|^2=1024$ possible imitation processes, but not all of these are important for the evolution of the population. In an early phase the naive strategies die out and the always defecting strategies become more popular. After this initial phase of evolution the nice and retaliating strategies take over. The most important transitions are shown in figure \ref{fig:chart}.
We can see that some of the always defecting strategies may directly be turned into GT or Pavlov by imitating one of the nice retaliating strategies, however none of the always defecting strategies can be turned into TFT or C$|$CCDC by imitating one of the nice and retaliating strategies. In general, it is more difficult for any always defecting strategies to be turned into TFT or C$|$CCDC than to be turned into Pavlov or GT. While this observation cannot explain the fate of all different strategies, it serves to explain -- together with the initial rise of always defecting strategies -- the lower equilibrium densities of the C$|$CCDC and TFT strategy when pIR is used. From this observation we may also conjecture that performance or fitness do not assure survival if the strategy cannot be easily learned by an important group of strategies. Thus the ``learnability'' of a strategy becomes an important factor that may have strong influence on the competitiveness. The Pavlov strategy is not successful in the high noise setting because it scores considerably lower than GT or TFT in a population with a large fraction of always defecting strategies.
\section{Conclusion \label{sec:conclusion}}
We have shown that the partial Imiation Rule, an alternative to the common imitative behaviour for two player games, can be described by an approximative mean value equation. Dynamics and stationary properties are in general subject to many parameters such as payoff parameters, noise and initial conditions. Our investigation is by no means exhaustive but the approximate mean value equation predicts that the evolution of well mixed populations depends heavily on the employed imitation rule. It is therefore important to discuss the imitation behaviour whenever meta-strategies, that are more complex than simply cooperate and defect, are being used.

The idea of using tIR or pIR (or any other imitation rule) is a question of the model one is trying to investigate. If we assume that offsprings are created from generation to generation it is meaningful to assume that they will be using the same strategy as the parent. This scenario is equivalent to using tIR. If on the other hands the players are surviving over several generations and are using imitation then they should adapt their strategies via a learning process (as for example pIR) rather than complete imitation or tIR.

The discussion about the imitative behaviour can also be taken to the spatial variant of the prisoner dilemma game, especially because in the spatial variant with stationary topology it is more natural to use imitation rather than reproduction. In the spatial variant one can argue that the imitator should, at least to some extent, have access to information from the games played by the role model with other players. We leave these topics to future work.
\begin{acknowledgments}
K.Y. Szeto acknowledges the support of CERG grant 602507. The authors thank Wenjin Chen for valuable discussions.
\end{acknowledgments}
\appendix
\section{Macroscopic dynamics \label{sec:app}}
In this section we determine the macroscopic dynamics of pIR from the microscopic interactions between the players. Our approach is mainly based on \cite{Szabo07}. For simplicity we consider only players on a fully connected network (every player interacts with every other player and himself). The transition rate of strategy $i$ to another strategy $j$ is given by

\begin{equation}
w(i \rightarrow j) = \rho_j g(U_j-U_i) \label{eq:transa} \, ,
\end{equation}

where $\rho_j$ is the density of $j$-strategists, the smoothing function has been given previously and it is understood that the payoffs depend on the densities $\rho_1, \rho_2, ..., \rho_|M_n|$, where $|M_n|$ is the number of possible strategies. The \textit{approximate mean value equation} for a strategy $i$ in the general case is

\begin{equation}
\frac{d \rho_i}{dt}= \sum\limits_{j\neq i} \left[ \rho_j w(j \rightarrow i) - \rho_i w(i \rightarrow j) \right] \label{eq:meana}
\end{equation}

Inserting equation (\ref{eq:transa}) into equation (\ref{eq:meana}) yields a non-linear differential equation for the time derivative of every strategy\footnote{Note that if proportional imitation is used rather than smoothed imitation this procedure simply yields the replicator equation.}.

However in the case of pIR this does in general not yield useful results as it does not take any account of the fact that an $i$-strategist who imitates a $j$-strategist will in general not become a $j$-strategist himself. In order for us to take account of this we need to find a correct expression for the term $w(i \rightarrow j)$ in equation (\ref{eq:meana}). We define the following mapping
\begin{equation}
p : M \times M \times M  \rightarrow [ 0, 1 ] \,.
\end{equation}
The value $p(k,j,i)$ is the probability that a $k$-strategist becomes an $i$-strategist by imitating a $j$-strategist. The transtition rate for $k$-strategists becoming $i$-strategists by imitating $j$ strategists is
\begin{equation}
w(k\xrightarrow{j} i) = \rho_j g(U_j-U_k)p(k,j,i) \, .
\end{equation}
The the total transition rate for $k$-strategists migrating to strategy $i$ is thus
\begin{equation}
w(k\rightarrow i) =\sum_j \rho_j g(U_j-U_k)p(k,j,i) \, ,
\end{equation}
and the total fraction that migrates to strategy $i$ is
\begin{eqnarray}
f_+(\rightarrow i) &=& \sum_{k\neq i} \rho_k \sum_j \rho_j g(U_j-U_k)p(k,j,i) \nonumber \\
&=& \sum_j \rho_j \sum_{k\neq i} \rho_k  g(U_j-U_k)p(k,j,i)
\end{eqnarray}
In contrast to the case of tIR a second summation over appears here. Under pIR children strategies can be different from both of the parent strategies. Therefore we need to consider all the strategies (except strategy $i$) as imitator when considering the migration to strategy $i$ when the role model uses strategy $j$. This leads immediately to interesting phenomena as for example the possible rebirth of extinct strategies. Next we define a transition rate for $i$-strategists migrating to another strategy by imitating strategy $j$:
\begin{equation}
w(i \xrightarrow{j}) = \rho_j g(U_j-U_i) [ 1 - p(i,j,i) ] \, 
\end{equation}
where the term in brackets takes care of the important case where the $i$-strategist learns nothing new by imitating a $j$-strategist and therefore keeps his previous strategy. The fraction migrating away from strategy $i$ is thus
\begin{equation}
f_-(i\rightarrow) = \rho_i \sum_j \rho_j g(U_j-U_i) [ 1 - p(i,j,i) ]
\end{equation}

The approximate mean value equation then becomes
\begin{eqnarray}
\frac{d \rho_i}{dt}= &&
f_+(\rightarrow i) - f_-(i\rightarrow) \nonumber \\
= &&\sum\limits_{j}   \rho_j \sum\limits_{k\neq i}  \rho_k g(U_j-U_k) p(k,j,i) \nonumber \\
 &-&\rho_i \sum\limits_{j} \rho_j g(U_j-U_i) [1- p(i,j,i)] \,. 
\end{eqnarray} 
The restrictions on the summation over $k$ and the factor in brackets are actually not necessary. We rewrite
\begin{eqnarray}
\frac{d \rho_i}{dt} &=& \phantom{-} \sum\limits_{j} \rho_j  \sum\limits_{k}  \rho_k  g(U_j-U_k)p (k,j,i) \nonumber \\
 && - \sum\limits_j \rho_i \rho_j g(U_j-U_i) p(i,j,i) \nonumber \\
 && - \sum\limits_{j} \rho_i \rho_j g(U_j-U_i) \nonumber \\ 
 && + \sum\limits_j \rho_i \rho_j g(U_j-U_i) p(i,j,i) \,
\end{eqnarray}
and see that two terms on the second and forth line cancel. Finally we obtain the general approximate mean value equation:
\begin{equation}
\frac{d \rho_i}{dt}= \sum\limits_{j} \rho_j \sum\limits_{k}   \rho_k g(U_j-U_k) p(k,j,i) -  \rho_i \sum\limits_j \rho_j g(U_j-U_i)  \,. \label{eq:mean_pIRa}
\end{equation}  
By specifying the values of $p$ we choose the imitation rule. For tIR we simply have
\begin{equation}
p_{\textrm{tIR}} (k,j,i) = \delta_{ij} \,,
\end{equation}
where $\delta$ is the Kronecker delta and the equation reduces to the approximate mean value equation in \cite{Szabo07}. For pIR we have
\begin{eqnarray}
p_{\textrm{pIR}}(k,j,i) &=& \left \lbrace \begin{array}{l}1\;\;\textrm{if $k$-strategist imitating} \\
\textrm{\phantom {1}\;\;$j$-strategist with pIR}\\
\textrm{\phantom {1}\;\;becomes $i$-strategist} \\ 
0\;\;\textrm{otherwise} \end{array} \right. \label{eq:deltaa}.
\end{eqnarray}
\vspace*{1pt}
\bibliography{../refs/pd_refs}
%
\end{document}